\documentclass[twocolumn,a4paper,10pt,twoside]{article}
\pdfoutput=1

\usepackage[T1]{fontenc}
\usepackage[utf8]{inputenc}
\usepackage[british]{babel}
\usepackage{amsmath}
\usepackage{textcomp}
\usepackage{amssymb}
\usepackage{url}
\usepackage{hyperref}
\usepackage{cleveref}
\usepackage{booktabs}
\usepackage[noprefix]{nomencl}	
\usepackage{stmaryrd}
\usepackage{xspace}
\renewcommand{\cite}{\citep}
\usepackage{fancyhdr}
\usepackage{natbib}
\usepackage{graphicx}

\usepackage[activate={true,nocompatibility},final,kerning=true,tracking=true,spacing=true]{microtype}
\microtypecontext{spacing=nonfrench}

\newcommand{\x}{$\times$}
\newcommand{\eg}{e.g.\xspace}
\newcommand{\ie}{i.e.\xspace}

\renewcommand{\v}[1]{\ensuremath{\mathbf{#1}}} 
\newcommand{\gv}[1]{\ensuremath{\mbox{\boldmath$ #1 $}}}  
\newcommand{\pd}[2]{\frac{\partial #1}{\partial #2}} 
\newcommand{\grad}[1]{\gv{\nabla} #1} 
\renewcommand{\div}[1]{\gv{\nabla} \cdot #1} 
\let\baraccent=\= 
\renewcommand{\=}[1]{\stackrel{#1}{=}} 
\newcommand{\nomunit}[1]{\renewcommand{\nomentryend}{\hspace*{\fill}#1}}

\title{Extending a serial 3D two-phase CFD code to parallel execution
 over MPI by using the PETSc library for domain decomposition}
 \author{{\AA}smund Ervik, Svend Tollak Munkejord and Bernhard M\"{u}ller}
 \date{\today\\[12pt]This paper has been submitted to the CFD 2014 conference.}

\crefname{chapter}{Chapter}{Chapters}
\crefname{section}{Section}{Sections}
\crefname{appendix}{Appendix}{Appendices}
\crefname{subsection}{Section}{Sections}
\crefname{subsubsection}{Section}{Sections}
\crefname{equation}{Eq.}{Eqs.}
\crefname{figure}{Fig.}{Figs.}
\crefname{table}{Table}{Tables}
\crefname{subfigure}{Fig.}{Figs.}
\crefname{listing}{Listing}{Listings}

\makenomenclature

\begin{document}
\maketitle 

\vspace{-15pt}

\abstract{

\setlength{\parindent}{15pt}

To leverage the last two decades' transition in High-Performance Computing
  (HPC) towards clusters of compute nodes bound together with fast
  interconnects, a modern scalable CFD code must be able to efficiently
  distribute work amongst several nodes using the Message Passing Interface
  (MPI). MPI can enable very large simulations running on very large
  clusters, but it is necessary that the bulk of the CFD code be written with MPI
  in mind, an obstacle to parallelizing an existing serial code. 

  In this work we present the results of extending an existing two-phase 3D
  Navier-Stokes solver, which was completely serial, to a parallel execution
  model using MPI. The 3D Navier-Stokes equations for two immiscible
  incompressible fluids are solved by the continuum surface force 
  method, while the location of the interface is determined by the level-set
  method. 

  We employ the Portable Extensible Toolkit for Scientific Computing (PETSc)
  for domain decomposition (DD)
  in a framework where only a fraction of the code needs to be
  altered. We study the strong and weak
  scaling of the resulting code. Cases are studied that are relevant to the
  fundamental understanding of oil/water separation in electrocoalescers.
}

\vspace{-10pt}

\printnomenclature[0.0cm]  
\vskip .1em

\section{Introduction}
In 1965 Gordon Moore famously predicted that transistor density (and hence
computing power for a given chip) would double each year in the foreseeable
future \cite{moore1965}. Dubbed Moore's law, this trend continued to hold for 
roughly 40 years and meant that life was easy for people needing greater and 
greater computational power. While serious High-Performance Computing (HPC) was
dominated in most of this period by vector machines like the seminal Cray 1,
by the mid-1990s clusters of many interconnected scalar
CPUs had become a cheaper solution, leading to the industry-wide adoption of
distributed memory architectures. 

Around 2005 Moore's law finally started hitting a barrier when 
the high heat production of chips and, somewhat later, the diffraction 
limits for photolitography began forcing chip makers to alter their ways. Two
complementary solutions were introduced, namely shared-memory architectures
(multi-core CPUs) and vector instruction sets (SSE, AVX, FMA)\footnote{SSE:
Streaming SIMD Extensions. AVX: Advanced Vector Extensions. FMA: Fused
Multiply-Add.}. 
Both solutions were adopted in HPC, leading to hybrid 
shared-memory/distributed-memory systems. In the last five years accelerator
technologies (GPGPU, MIC)\footnote{GPGPU: General-Purpose Graphics Processing
Unit. MIC: Many Integrated Core.} have furthered the return to vector processing, so HPC
has in a sense come full circle. All in all this gives a very heterogeneous
environment for HPC where the onus is on the application programmer to ensure
that his/her code can make the most of the available resources. 

In contemporary numerical codes, omitting here the use of accelerators, the two
main programming paradigms for leveraging parallelism are OpenMP and MPI. OpenMP
takes advantage of shared-memory architectures, while MPI can use
distributed-memory architectures. On current systems, OpenMP can scale from 1 to
32 cores, while MPI can scale to thousands and even millions of cores. This
means that MPI is the paradigm of choice for HPC, possibly in combination with
OpenMP used by each MPI process. 

We will use the following nomenclature when discussing parallelism:
a ``process'' is one MPI rank which is executing code. A CPU has several
``cores'', each of which may execute a process. The CPUs are located on
``nodes'', \eg a desktop computer or a blade in a cluster. Typical cluster nodes
have 2 (or more) CPUs, each having a separate ``socket'' connecting the CPU to
the memory (RAM). Each socket has one communication channel to memory shared by
all cores on this socket. Many nodes can communicate over the ``interconnect'',
which should preferrably be very fast and have very low latency.

This paper will focus on the use of MPI to port an existing serial
implementation of a 2D/3D incompressible Navier-Stokes solver. This code can
simulate two-phase flows relevant \eg for the fundamental understanding of
oil/water separation, but for 3D cases the runtime is very long (weeks and
months). The majority of this runtime is due to the solution of a Poisson
equation for the pressure, and state of the art algorithms for this problem are
bound by the memory bandwidth rather than CPU speed. This makes OpenMP a poor
solution in this case and leaves MPI as the necessary paradigm for parallelism. We
will employ the PETSc library, specifically the DMDA component, to do domain
decomposition. The solution of the Poisson equation is also done using PETSc
routines. We establish a framework in Fortran where it is possible to reuse the
existing serial code. 

The rest of this paper is organized as follows: in the next section, the basic
equations are established, after which the numerical methods are presented.
Then we describe the framework and the specific changes that were needed to port
the serial code. Computations performed with the resulting code are
discussed and we study the strong and weak scaling on several architectures.
Finally some closing remarks are given.

\section{Model Description}
The equations that govern the two-phase flow system under consideration are the
incompressible Navier-Stokes equations:
\begin{align}
  \div\v{u} &= 0 \label{eq:ns-divfree} \\
	\pd{\v{u}}{t}+(\v{u}\cdot\grad)\v{u} &= - \frac{\grad{p}}{\rho}
  + \frac{\mu}{\rho}\grad^2\v{u} +
	\v{f}
	\label{eq:ns}
\end{align}%
\nomenclature{$\v{u}(\v{x})$}{Velocity field of a fluid.\nomunit{m/s}}%
\nomenclature{$\mu$}{Dynamic viscosity of a fluid.\nomunit{Pa$\cdot$s}}%
\nomenclature{$\nu$}{Kinematic viscosity of a fluid.\nomunit{m$^2$/s}}%
\nomenclature{$\rho$}{Density of a fluid.\nomunit{kg/m$^{3}$}}%
\nomenclature{$p(\v{x})$}{Pressure of a fluid.\nomunit{Pa}}%
\nomenclature{$\v{f}$}{External acceleration.\nomunit{m/s$^2$}}%

These equations are valid for single-phase flow. To extend this
formulation to two-phase flow we keep these equations in each of the two
phases, where the densities and viscosities are constant in each phase. We 
will restrict ourselves to laminar flow, as we are interested in situations with
Reynolds numbers $\rm{Re} \sim \mathcal{O}(1)$. 

Across the interface between the fluids, a jump in the normal component of the
traction vector will arise due to the surface tension $\sigma$, and this jump together with
effects of the jump in density and viscosity must be added to our equations.  We
introduce these effects using the continuum surface force method (CSF)
\cite{brackbill1992}. The location of the interface is captured using the
level-set method (LSM) \cite{osher1988,osher2001}, see \citet{ervik2014} for
a detailed description, we provide only a short outline here.
\nomenclature[2]{$\sigma$}{Coefficient of surface tension.\nomunit{N/m}}%
\nomenclature[2]{$\kappa$}{Curvature of the interface.\nomunit{1/m}}%
\nomenclature[4]{Re}{Reynolds number.}

The level-set method is a method for capturing the location of an interface. It
is widely used not just for multi-phase fluid flow but also in other contexts
where an interface separates two regions. The interface is represented by
a level-set function $\phi(\v{x})$ which is equal to the signed distance to the
interface. In other words, the interface is given by the zero level set
$\{\v{x}\, |\, \phi(\v{x}) = 0\}$, hence the name. Rather than advecting the interface location, one
advects the function $\phi(\v{x})$ directly according to the transport equation
\begin{equation}
  \pd{\phi}{t} =  -\v{u}\cdot\grad\phi \rm{}
  \label{eq:advection}
\end{equation}
giving an implicit formulation that automatically handles changes in interface
topology. 

\begin{figure}[!htb]
  \centering
  \includegraphics[width=\linewidth]{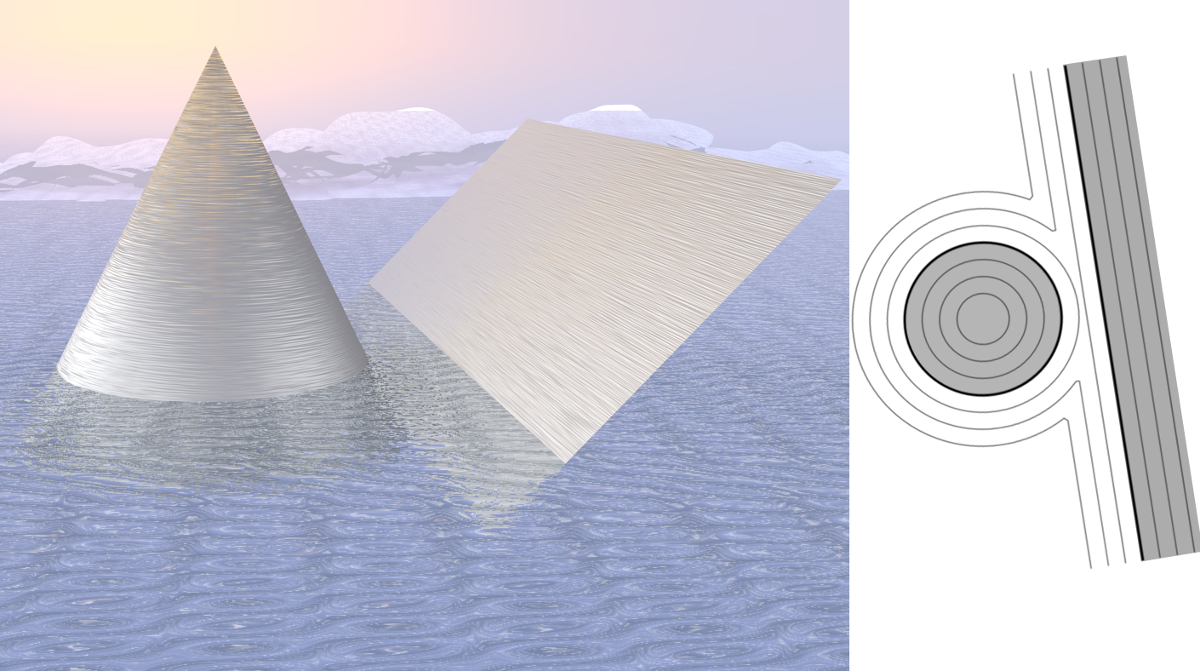}
  \caption{Illustration of the level-set method. Right: in 2D, a fluid drop 
    (dark gray) seen next to a fluid film (dark gray), both immersed in a 
    different fluid (white). Left: the signed-distance function representing
    these two fluid bodies, the drop and the film.}
  \label{fig:level-set}
\end{figure}

The level-set method can be visualized as
in \cref{fig:level-set} for a 2D fluid flow with a drop next to a film, seen on
the right-hand side in this figure as gray shapes. The distance is shown as
isocontour lines superimposed on these shapes. On the left-hand side the
level-set function is shown visualized in 3D as surfaces where the height above
water corresponds to the signed distance. The analogy to a map describing an
island rising out of the water is quite striking, except that the roles of
``reality'' and ``tool for description'' have been reversed.

When the location of the interface is known, the curvature $\kappa$ can be
calculated from $\phi$, and together with $\sigma$ this gives the surface
tension force. In the CSF method this force is incorporated as a volume-force
term which is non-zero only in a thin band around the interface. This thin band
is produced by smearing out the delta function, making the force term continuous
For such a smeared-out delta function, we can compute the volume-force term at
a point $\v{x}$ close to the interface as
\begin{equation}
	\v{f}_{\rm{s}}(\v{x},t) =
	\int_{\Gamma}\v{f}_{\rm{sfd}}(\v{s},t)\delta(\v{x}-\v{x}_{I}(s))d\v{s}\rm{,}
	\label{eq:surface-force}
\end{equation}
where $\v{f}_{\rm{sfd}}$ is a surface-force density 
and $\v{x}_{I}(s)$ is a parametrization of the interface. The surface-force
density is such that the integral of $\v{f}_{\rm{s}}(\v{x},t)$ across the
(smeared-out) interface approximates the surface tension force, see
\citet{brackbill1992} for details. Note that in
the level-set context it is not necessary to parametrize the interface
since $\phi(\v{x})$ stores the distance to the interface, so we have
$\v{x}-\v{x}_{I}(s) = \phi(\v{x})$ as long as $\phi(\v{x})$ is a signed distance
function. There are several ways to smear out the delta function, we follow 
\citet[Eq. 1.23]{osher2003},
\begin{equation}
  \delta(x) = 
  \begin{cases}
    0 & \rm{if} \; |\phi| > \epsilon \\
    \frac{1}{2\epsilon}\left(1
    + \cos\left(\frac{\pi\phi}{\epsilon}\right)\,\right) & \rm{else}
  \end{cases}
  \label{eq:deltafunc}
\end{equation}
where $\epsilon = 1.5 \Delta x$ is employed. This one-dimensional delta function
is composed into the three-dimensional version by taking $\delta(\v{x})
= \delta(x)\delta(y)\delta(z)$.

This formulation leads to a source term which incorporates the effects of
surface tension. It is also necessary to smear out the viscosity and
density differences across the interface in order to be consistent with the above
formulation. A smeared-out Heaviside function $H(\v{x})$ is used to accomplish 
this, given by \citet[Eq. 1.22]{osher2003} as
\begin{equation}
  H(x) = 
  \begin{cases}
    0 & \rm{if} \; \phi < -\epsilon \\
    \frac{1}{2}\left(1
    + \frac{\phi}{\epsilon} +  \frac{1}{\pi}\sin\left(\frac{\pi\phi}{\epsilon}\right)\,\right)
    & \rm{if} \; -\epsilon < \phi < \epsilon \\
    1 & \rm{if} \; \phi > \epsilon \\
  \end{cases}
  \label{eq:heaviside}
\end{equation}

\section{Numerical methods}
To discretize the Navier-Stokes equations and the equations for the level-set
method we employ finite difference methods, specifically
WENO \cite{liu1994} for the convective terms and central differences for the
viscous terms in \cref{eq:ns}, and WENO also for \cref{eq:advection}. 
The time integration is done with an explicit second-order
Runge-Kutta method (SSPRK (2,2) in the terminology of \citet{gottlieb2009}) for
both \cref{eq:ns} and \cref{eq:advection}. 

The
grid is a structured rectangular uniform staggered grid.  A staggered grid is
employed to avoid checkerboarding of the pressure field; this means that the
pressure and other scalars ``live'' at cell centers, while the velocities
``live'' at the cell faces. To be more precise, if we have a pressure at one
point $p_{i,j,k}$, the velocities $(u, v, w)$ around this point are
$u_{i\pm1/2,j,k}, v_{i,j\pm1/2,k}, w_{i,j,k\pm1/2}$ located at the 6 cell faces.
In the actual code we store the velocity values for
$u_{i+1/2,j,k},\,v_{i,j+1/2,k},\,w_{i,j,k+1/2}$ at the index \texttt{(i,j,k)}
even though these values are not physically colocated. 

The major problem when solving \cref{eq:ns-divfree,eq:ns} is that this is not
a set of PDEs, it is a differential-algebraic equation (DAE) with a Hessenberg
index of two. In other words, even though we have four equations (\cref{eq:ns} is three
equations) and four unknowns ($u,v,w,p$), \cref{eq:ns-divfree} cannot be used
directly to find $p$. 
The first solution to this conundrum was presented by 
\citet{chorin1968}.
This method can be understood as calculating an approximate velocity
field $\v{u}^*$ which does not satisfy \cref{eq:ns-divfree}, and subsequently
projecting this velocity field onto the manifold of vector fields satisfying
\cref{eq:ns-divfree}. For this reason, the method is often called Chorin's
projection method or simply \emph{the} projection method. 
It consists of these three steps, where we calculate three
quantities successively, namely $\v{u}^*,p_{n+1},\v{u}_{n+1}$:
\begin{align}
  \label{eq:intermed}
  \frac{\v{u}^*-\v{u}_n}{\Delta t} &= -\left( \v{u}_n\cdot\grad \right)\v{u}_n
  + \nu\grad^2\v{u}_n\\
  \label{eq:poisson}
  \grad^2 p_{n+1} &= \frac{\rho}{\Delta t}\div\v{u}^* \\
  \v{u}_{n+1} &= \v{u}^* - \frac{\Delta t}{\rho}\grad p_{n+1}
  \label{eq:chorin}
\end{align}
\nomenclature[4]{$n$}{Time step index.}

The pressure Poisson equation (\ref{eq:poisson}) that arises here is elliptic,
so the numerical solution is very time consuming and a vast amount of research
has gone into developing fast solvers. For two-phase flows with high density
differences, the condition number of the matrix that results when
\cref{eq:poisson} is discretized will make matters even worse than for the
single-phase problem \cite{duffy}. This matrix is very large even in sparse
storage formats, for a 256$^3$ grid it has 117 million non-zero elements.
The current state-of-the-art consists in
combining a (geometric or algebraic) multigrid preconditioner with a conjugate
gradient method (often BiCGStab) for solving the resulting sparse linear system.
Our experience with 2D axisymmetric simulations suggests that the Bi-Conjugate Gradient Stabilized method
\cite{vandervorst1992} with the BoomerAMG preconditioner \cite{henson2000} is
an optimal choice. For the simulations performed here, however,
the straigth-forward successive over-relaxation (SOR) preconditioner turned 
out to be faster than BoomerAMG. This has not been investigated in greater
detail. We employ the PETSc and Hypre libraries for these methods
\cite{petsc,hypre}.

We note
also that the boundary conditions for \cref{eq:poisson} are of pure Neumann
type (unless \eg an outlet pressure is specified), which results in a singular
matrix. These boundary conditions arise from the projection method and are not
physical. The common ``engineering'' approach of fixing the singularity, simply
fixing the pressure at some point in the domain, is not a very good approach as it
may pollute the spectrum of the preconditioner. Instead, projecting the
discretized singular equation into the orthogonal complement of the null space 
of the singular matrix seems to be a good solution \cite{Zhuang2001}. In other
words, for $Ax=b$, we construct the Krylov operator $K=(\mathbb{I}-\mathbb{N})P^{-1}A$ such that
${b,Kb,K^2b,...}$ is orthogonal to the null space $\mathbb{N}$. Here
$\mathbb{I}$ is the identity matrix, so $(\mathbb{I}-\mathbb{N})P^{-1}$ is the
desired projection. In the PETSc library that we
employ here \cite{petsc}, this is achieved using the \texttt{KSPSetNullSpace()} 
routine.

\section{Parallelization}
The starting point for the parallelization was an in-house code consisting of
a 2D/3D Navier-Stokes solver and a multi-physics framework that enables the
simulation of two-phase flows with the possibility of applying electric fields,
and/or adding surface-active agents to the interface. The interface between the
two phases is captured using a level-set method, so interfaces with changing
topology such as two merging drops can be simulated.  The code has been
successfully applied to the study of both liquid-liquid \cite{teigen2010a} and
liquid-gas systems \cite{ervik2014}, but the long runtime has restricted its use
to 2D axisymmetric cases so far. 

The PETSc DMDA framework for domain decomposition was chosen as the main
methodology for parallelizing the code. Domain decomposition consists in
splitting the whole domain into subdomains which are each distributed to one
MPI node. Each node then has a computational domain with some internal cells
where the flow is computed, and some ghost cells which represent either boundary
conditions or values that belong to neighbouring domains. This means that
regular communication between the nodes is necessary such that all ghost cells
have correct values. Such a splitting is shown in \cref{fig:mfg-soln} below.
Neglecting for a moment the pressure Poisson equation, this
approach can scale well to millions of CPU cores, see \eg
\citet{SC2013GordonBell} for an example in compressible flow. 

By using the PETSc DMDA framework we can avoid the gritty details of domain
decomposition and MPI programming. At the initialization of the code, some
routines are called to set up three DMDAs which are objects that manage the
decomposition. Using these objects we input how large our computational
domain should be in terms of grid points, and the library decides an optimal
decomposition \emph{at runtime} depending on the number of MPI processes the code is
run with. We also specify the physical dimensions of our uniform grid, and the
library returns the physical dimensions for each subdomain.

This framework is very convenient, but one enhancement was made to further
facilitate the reuse of the serial code. In the standard PETSc framework, the 
local work arrays that represent the solution on a given subdomain and the 
values in the ghost cells are indexed using the global indices. The existing
code naturally expects indices that go from 1 to the maximum value
\texttt{imax}. In Fortran, the bounds of an array may be re-mapped when the
array is passed to a subroutine, and this feature was used to ensure that each
local work array had bounds as expected by the serial code. Thus we will use
\texttt{imax} as the final \texttt{i} index \emph{on each subdomain} in the
following.

With this enhancement, the only thing that had to be re-written in the original
code was the 
handling of the staggered grid for the velocity. In the formulation used here,
we have one less point for \eg $u$ in the $x$ direction, since these values are
located at the cell faces. In the serial code this is handled by not solving for
$u$ at the point \texttt{imax}. In the parallel version, $u$ at the point
\texttt{imax} should however be solved for on those processes that are not at
the actual boundary but where there is a neighbouring process in the
positive $x$-direction. 

Furthermore, this means that a communication step is also
necessary in the projection method. After the pressure has been calculated from
the Poisson equation, we calculate \eg the $x$-component of $\grad p$ at the
cell face corresponding to $u$ at \texttt{imax}. Numerically this is
\texttt{(p(imax+1) - p(imax))/dx}, so the ghost value at 
\texttt{imax+1} must be updated before this calculation for those subdomains
where \texttt{p(imax+1)} represents a pressure
value on another subdomain and not a boundary condition.

Returning to the pressure Poisson equation \cref{eq:poisson}, the elliptic nature of this
equation means that, in some sense, all
nodes must communicate during the solution. A further
reduction in speedup potential is due to the fact that the solvers for this equation are mostly 
bound by memory bandwidth, which is shared amongst all cores on a modern CPU.
These limits imply that we must lower our expectations somewhat in comparison
with the impressive results mentioned earlier for compressible flows.

In the DMDA framework, the Poisson equation is set up such that each process
computes its own portion of the matrix and right-hand side vector. This is the
only scalable way of solving it, even when sparse storage formats are used.

\section{Results}
\subsection{Manufactured solution case}
\begin{figure}[!htb]
  \centering
  \includegraphics[width=\linewidth]{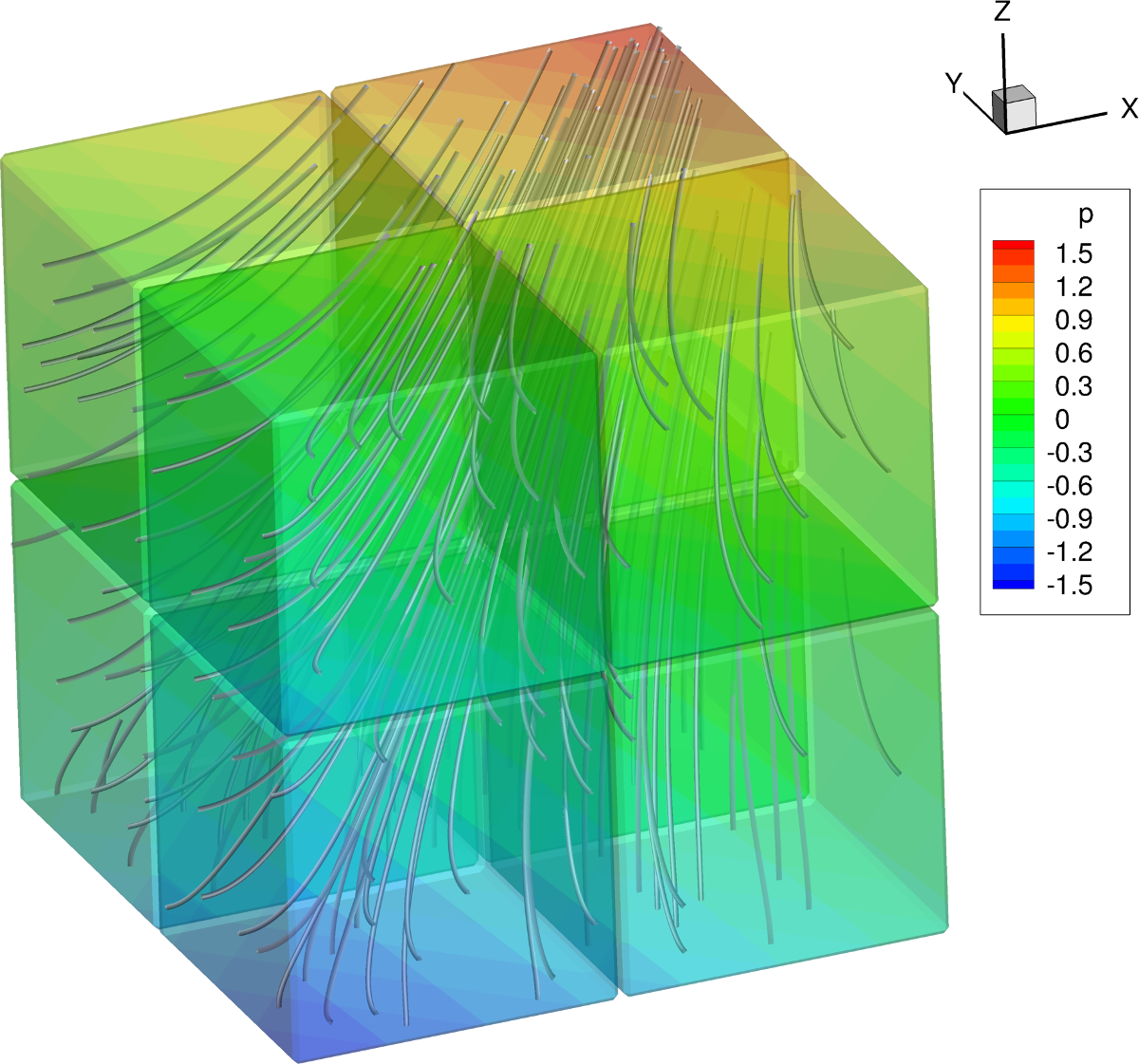}
  \caption{The computed solution after 0.031 s (100 time steps) on a 128$^3$ grid run on
  8 processors. The blocks show the decomposition of the domain, the pressure
field is shown superimposed on these blocks, and the streamlines illustrate the
flow.}
  \label{fig:mfg-soln}
\end{figure}

After the code had been parallelized it was tested 
using a manufactured solution \cite{roache2002} inspired by that used in \citet{john2006}. The
debugging tool Valgrind \cite{2007valgrind,valgrind} was used in the memory checking mode to
ensure that the code does not \eg make use of uninitialized values, a common
programming error. When all such errors were fixed, the
code was used to solve the single-phase Navier-Stokes equations with the
following exact solution used as an initial -- boundary value problem on a 
(1.0 m)$^3$ domain, where the origin is in the lower left front corner (cf.\
\cref{fig:mfg-soln}).
\begin{equation}
\begin{aligned}
  u &= t^3 y z \\
  v &= t^2 x z\\
  w &= t x y\\
  p &= x + y + z - 1.5 \\
\end{aligned}
\label{eq:mfg-soln}
\end{equation}

Insertion into \cref{eq:ns-divfree} confirms that this solution is divergence free, 
and the resulting body force can be
computed by inserting \cref{eq:mfg-soln} in \cref{eq:ns}. In order to minimize the risk of
human error, this was done symbolically using
Maple, the resulting expression was copied into the Fortran code and regular
expressions were used to convert Maple syntax into Fortran. A plot of the
computed solution is shown in \cref{fig:mfg-soln}. Here the velocity streamlines
are shown together with the pressure field which has been superimposed on blocks
representing the parallel decomposition.

\subsection{Convergence}
Using the manufactured solution in \cref{eq:mfg-soln}, the convergence under
grid- and time step refinement, as well as the strong and weak scaling, was
tested on the Kongull cluster at NTNU. This cluster has dual-socket nodes with
Intel Xeon E5-2670 8-core CPUs and a 1 Gb/s Ethernet interconnect. The STREAM
benchmark \cite{stream,McCalpin1995} was run on one core and gave an effective memory
bandwidth of 9800 MB/s for the Triad test\footnote{The Triad test consists of
  repeated computations of the operation \texttt{a(i)=b(i)+q*c(i)} where
  \texttt{q} is constant and \texttt{i} is incremented.}. 

  \begin{figure}[!htb]
  \centering
  \includegraphics[width=\linewidth]{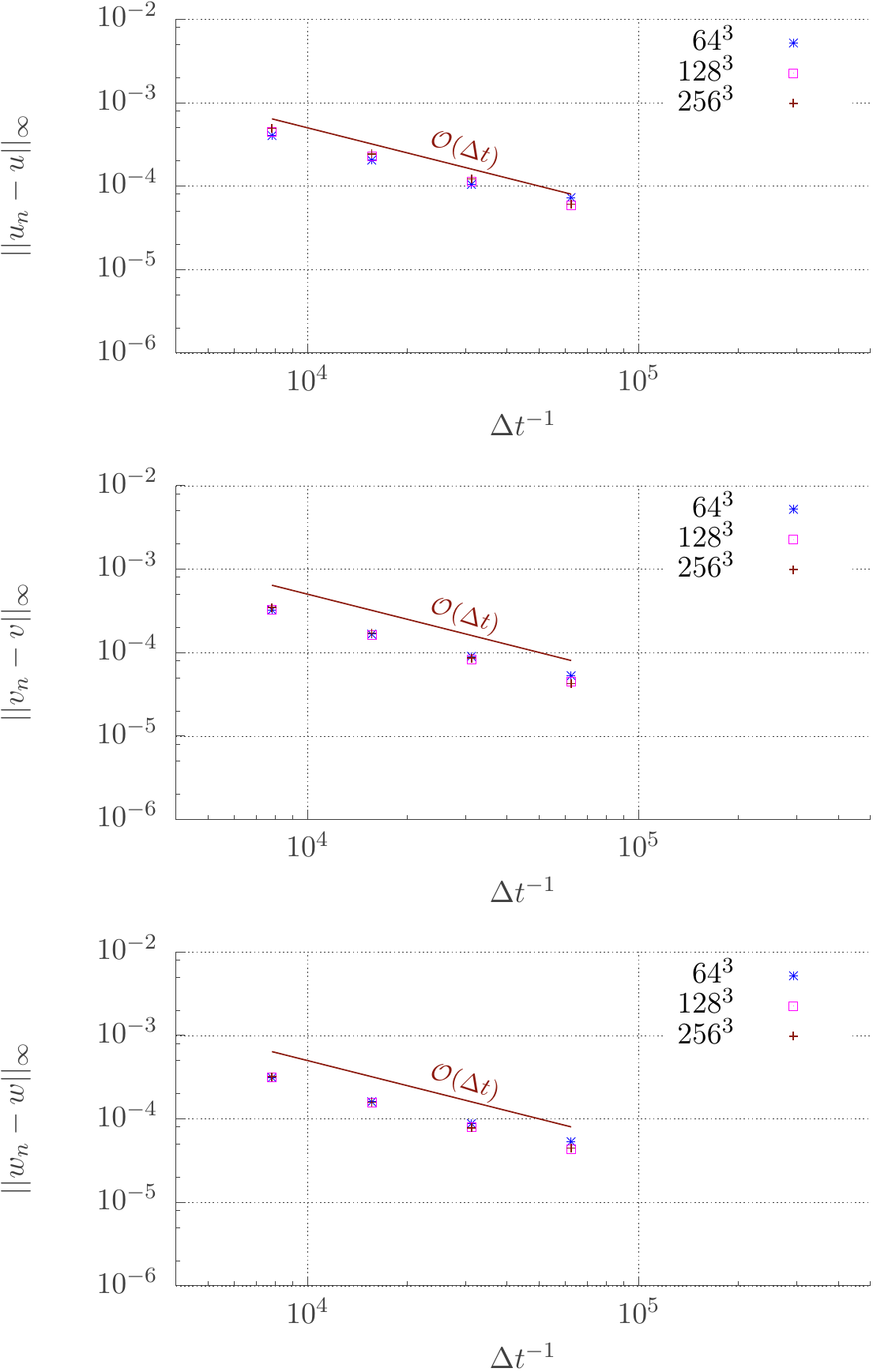}
  \caption{Time step and grid refinement on 32 processes. Top to
  bottom: $u$, $v$ and $w$ velocities. The maximum error of the solution, 
  \eg $||u_n - u ||_\infty$ is plotted against the inverse of the time step. 
  Here $u_n$ denotes the numerical solution at time step $n$ while $u$ denotes
  the exact solution at this time.}
  \label{fig:refinement}
\end{figure}

To test the grid- and time step refinement, a base case was selected with
a 256$^3$ grid, giving a grid spacing \texttt{dx} of $3.91\cdot10^{-3}$ m,
the CFL condition following \citet{kang2000} with a CFL number of 0.5 then giving
a time step of $1.28\cdot10^{-4}$ s. This case was solved for 100 time steps, as
were solutions on coarser grids 128$^3$ and 64$^3$ computed with the same time
step. All simulations were run on 32 processes (8 nodes with 4 processes each).
Subsequently, the same cases were run but with $1/2$, $1/4$ and $1/8$ the time
step using 200, 400 and 800 time steps, respectively. The results are shown in
\cref{fig:refinement}.

It is seen that the convergence behaviour is as expected.  First of all, the temporal order is
1 (not 2) due to an irreducible splitting error from the projection method. This can be
overcome \eg by using the incremental pressure form (see \citet{guermond2006} for
a review of projection methods), but has not
been considered in this work. Second, the grid refinement does not influence the
error. This is due to the fact that the velocity field is linear in space, so
the error is completely dominated by the temporal order.

\subsection{Strong scaling}
\begin{figure}[!htb]
  \centering
  \includegraphics[width=\linewidth]{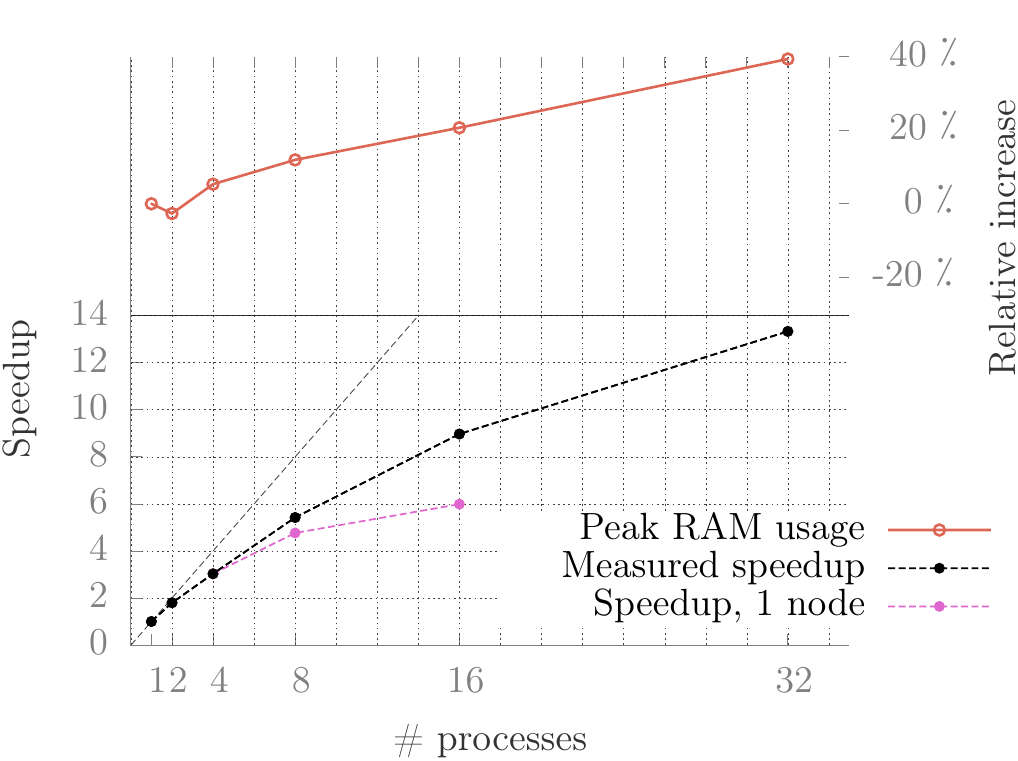}
  \caption{Strong scaling: with the left-hand-side $y$-axis, measured (black and magenta) and 
    the optimal (dotted gray) speedup plotted against the number of processes.
    With the right-hand-side $y$-axis, increase in memory use.}
  \label{fig:strong-scaling}
\end{figure}

To test the strong scaling of our code, \ie how simulating a given case speeds
up when more processes are used, a 128$^3$ grid was used giving a grid spacing
\texttt{dx} of $7.81\cdot10^{-3}$ m, the CFL condition giving a time step of
$3.10\cdot10^{-4}$ s. The solution was computed for 100 time steps.  Since the
Poisson solver performance should be bound by memory bandwidth, the test was
made using 2 processes per node (one per socket) and several nodes. The
resulting speedup relative to one process is shown in \cref{fig:strong-scaling}.
In this figure, the black points indicate the speedup compared to running on
one process. The scaling seen is quite good, but as expected lower than the
theoretical linear scaling. It is seen that the peak memory usage (orange)
increases slightly with more processes.

To investigate the hypothesis that using only 2 processes per node and several
nodes is better than using many processes on one node, we also tried running
with 8 and 16 processes on one node. These results are plotted in magenta in
\cref{fig:strong-scaling} and confirm the hypothesis.  We can conclude that even
on this particular cluster with a slow (by HPC standards) interconnect, the
added memory bandwidth afforded by using more nodes (thus more sockets)
outweighs the penalty of increased communication between nodes. This also
indicates that the results for 2 processes per node are bound by the
interconnect speed, such that the speedup would be closer to the optimal
(linear) scaling when run on a more tightly-coupled cluster.

\subsection{Weak scaling}
\begin{figure}[!htb]
  \centering
  \includegraphics[width=\linewidth]{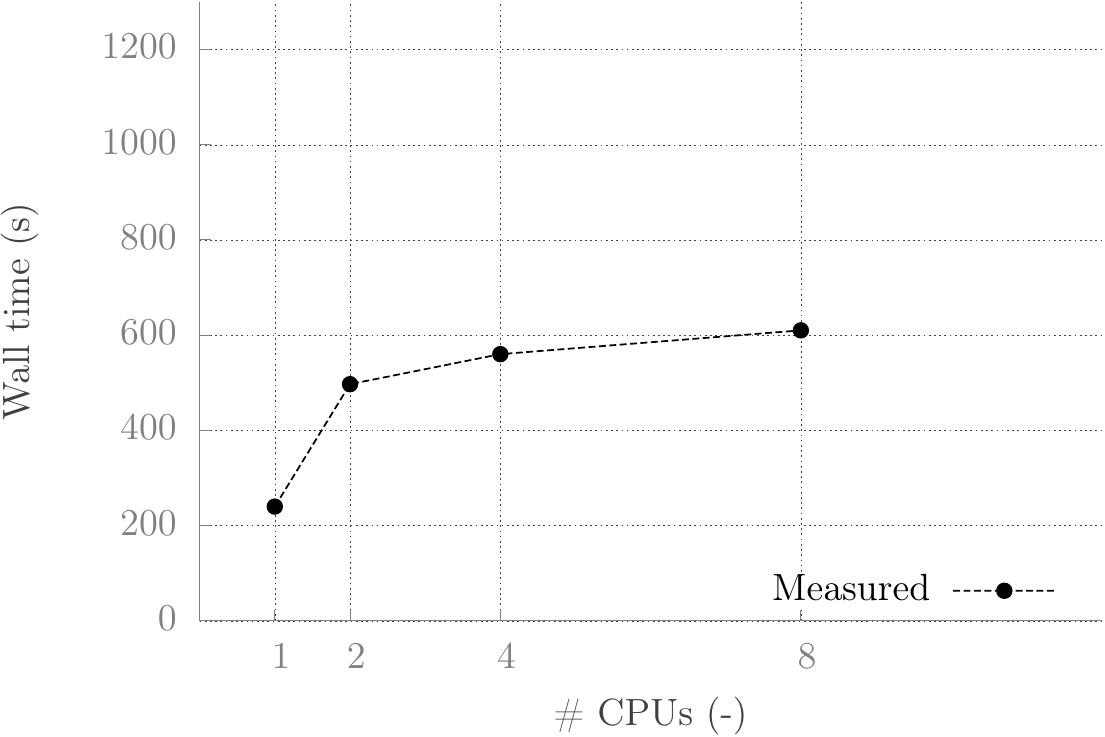}
  \caption{The weak scaling of the code as the number of processes and the
  number of grid points are both increased proportionally.}
  \label{fig:weak-scaling}
\end{figure}

The weak scaling of the code was also studied. The base case was the same
manufactured solution on a (0.5 m)$^3$ domain resolved with a 64$^3$ grid, run
on one process. Then a (0.5 m)$^2$\x(1.0 m) domain with a 64$^2$\x128 grid was
solved with two processes, a (0.5 m)\x(1.0 m)$^2$ domain with a 64\x128$^2$
grid was solved on 4 processes, etc. In this way, the number of grid points and
the number of processes are both increased proportionally. The equations were 
solved for 50 time steps, and the results are shown in \cref{fig:weak-scaling}.

As is seen in this figure, there is obviously a performance hit initially; the
perfect behaviour would be a flat line. This
is as expected. When going from 1 to 2 processes, we go from no communication to
overhead from communication. Furthermore, when going from 2 to 4 processes,
there is the added overhead of intra-node communication, as opposed to the case
with 2 processes where the communication is not over the network but over the
CPU bus. The weak scaling seen here is quite decent. 
One should also be aware that it is more difficult to ensure that 
cases are ``equivalently hard'' for weak scaling than for strong scaling
\cite{aagaard2013}.

\subsection{Two-phase results}
\begin{figure}[!htb]
  \centering
  \includegraphics[width=\linewidth]{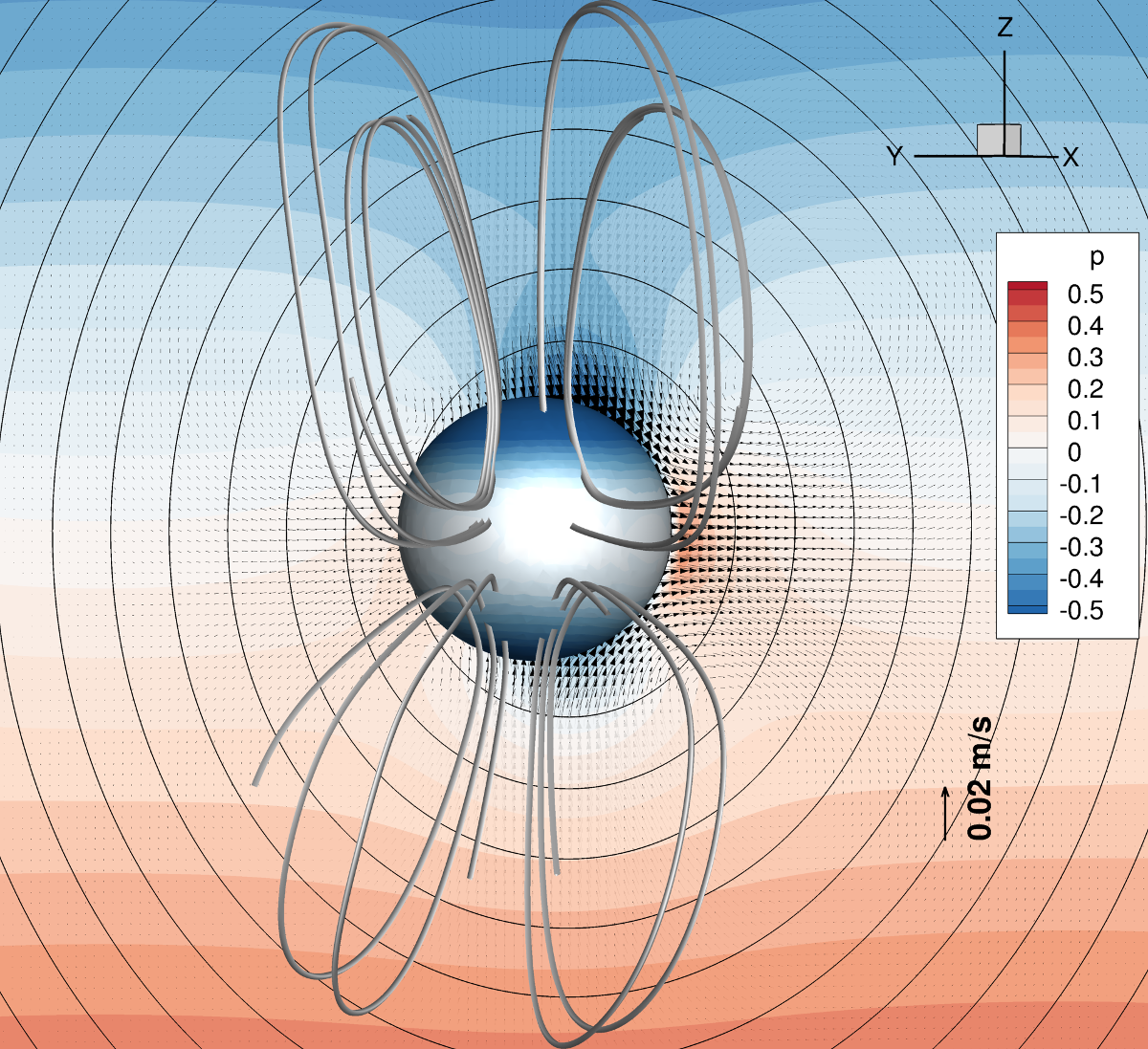}
  \caption{The falling drop after a short time ($0.01$ s). The
  pressure field is shown superimposed on the surface, and on the plane behind
  the drop. On this plane the velocity field and the level-set isocontours are
also shown. Streamlines indicate the velocity field.}
  \label{fig:drop}
\end{figure}

As an initial test of the two-phase capabilities of the parallelized code, the
CSF method was employed to simulate a 2 cm diameter drop with properties
$\rho_1=2$~kg/m$^3$, $\mu_1=0.01$~Pa~s falling through a bulk fluid with properties
$\rho_2=1$~kg/m$^3$, $\mu_2=0.05$~Pa~s. The interfacial tension was set to 
$\sigma=0.01$~N/m. The domain was (10 cm)$^3$ resolved by a (128)$^3$ grid, the
simulation was run on 8 processes for 33900 time steps up to $t=0.01$~s. The
drop has not yet achieved a substantial falling velocity, so the spurious
currents are quite visible. The result is shown in \cref{fig:drop}, where the
drop is shown with the pressure superimposed on the surface, streamlines
indicating the flow. A plane is shown intersecting the centre of the drop, on
this plane the pressure field, velocity field and level-set function contour
lines are shown. A reference vector is shown on the right.

Spurious currents is a well-known challenge with the CSF method,
and experience with the 2D serial code has led us to prefer the ghost-fluid
method (GFM) \cite{kang2000}, which is somewhat more complicated to implement.
This was not done within the scope of this paper. Nevertheless, this
demonstrates that the parallel code is capable of two-phase fluid simulations
with both density- and viscosity-jumps.

\section{Conclusions}
In this paper we have discussed the parallelization of an existing serial 3D
incompressible Navier-Stokes solver for two-phase flow. The PETSc DMDA domain
decomposition framework has been leveraged to apply MPI parallelism, enabling
the code to make use of modern HPC facilites. We have discussed the
alterations that were necessary for the serial code and established a framework
where these were as few as possible.

Based on this code, we have reported the strong and weak scalings for
a manufactured-solution case on a cluster with dual-socket nodes and 1 Gb/s
Ethernet interconnect. It is seen that the code scales rather well, but that one
should take care to maximize the number of sockets used, since the Poisson
solver is bound by memory bandwidth. If this code is run on a cluster
simultaneously with CPU-bound parallell codes (\eg using Monte Carlo methods),
sensible resource allocation would benefit from taking the available memory
bandwidth into account. Then it would not be optimal to allocate all cores on
$N$ nodes to this code, but rather \eg 50\% of the cores on $2N$ nodes, while
a CPU-bound code could effectively utilize the remaining 50\% of the cores. 

The speedup seen in the strong scaling test (13x faster on 32 processes) is
sub-linear but does not level-off.  Together with the possibility of running on
more tightly-coupled clusters where the behaviour should be closer to linear,
and using more than 32 cores, this will give a substantial speedup and reduce
the runtimes of weeks and months for the serial code to something
more managable, \ie a few days or less.

Initial tests demonstrate that the code is able simulate two-phase flow, but the
ghost-fluid method (GFM) should be used instead of the CSF method currently
employed, in order to minimize the spurious currents.

This effort has left us with a code that scales quite well and a framework where the
remaining multi-physics components can easily be introduced. In the end this
will enable future simulations of full 3D cases relevant for the fundamental
understanding of electrocoalescence.

\section{Acknowledgement}
We would like to thank Matthew Knepley, Barry Smith and Jed Brown of the PETSc project
for useful discussions around the framework employed here.

This work was funded by the project \emph{Fundamental understanding of
electrocoalescence in heavy crude oils} coordinated by SINTEF Energy Research.
The authors acknowledge the support from the Petromaks programme of the Research
Council of Norway (206976), Petrobras, Statoil and W\"{a}rtsil\"{a} Oil \& Gas Systems.

\bibliographystyle{CFD2014}
\bibliography{references}

\begin{thebibliography}{26}
\newcommand{\enquote}[1]{``#1''}
\providecommand{\natexlab}[1]{#1}
\providecommand{\url}[1]{\texttt{#1}}
\providecommand{\urlprefix}{URL }
\providecommand{\eprint}[2][]{\url{#2}}

\bibitem[{Aagaard \emph{et~al.}(2013)}]{aagaard2013}
AAGAARD, B.T. \emph{et~al.} (2013).
\newblock \enquote{A domain decomposition approach to implementing fault slip
  in finite-element models of quasi-static and dynamic crustal deformation}.
\newblock \emph{J. Geophys. Res.: Solid Earth}, \textbf{118(6)}, 3059--3079.

\bibitem[{Balay \emph{et~al.}(2014)}]{petsc}
BALAY, S. \emph{et~al.} (2014).
\newblock \enquote{{PETSc} {W}eb page}.
\newblock {\url{http://www.mcs.anl.gov/petsc}}.

\bibitem[{Brackbill \emph{et~al.}(1992)}]{brackbill1992}
BRACKBILL, J. \emph{et~al.} (1992).
\newblock \enquote{A continuum method for modeling surface tension}.
\newblock \emph{J. Comput. Phys.}, \textbf{100(2)}, 335--354.

\bibitem[{Chorin(1968)}]{chorin1968}
CHORIN, A.J. (1968).
\newblock \enquote{Numerical solution of the {N}avier-{S}tokes equations}.
\newblock \emph{Math. Comput.}, \textbf{22(104)}, 745--762.

\bibitem[{Duffy \emph{et~al.}(2002)}]{duffy}
DUFFY, A. \emph{et~al.} (2002).
\newblock \enquote{An improved variable density pressure projection solver for
  adaptive meshes}.
\newblock Unpublished. See \url{http://www.math.fsu.edu/~sussman/MGAMR.pdf}.

\bibitem[{Ervik \emph{et~al.}(2014)}]{ervik2014}
ERVIK, {\AA}. \emph{et~al.} (2014).
\newblock \enquote{A robust method for calculating interface curvature and
  normal vectors using an extracted local level set}.
\newblock \emph{J. Comput. Phys.}, \textbf{257}, 259--277.

\bibitem[{Gottlieb \emph{et~al.}(2009)}]{gottlieb2009}
GOTTLIEB, S. \emph{et~al.} (2009).
\newblock \enquote{High order strong stability preserving time
  discretizations}.
\newblock \emph{J. Sci. Comput.}, \textbf{38(3)}, 251--289.

\bibitem[{Guermond \emph{et~al.}(2006)}]{guermond2006}
GUERMOND, J. \emph{et~al.} (2006).
\newblock \enquote{An overview of projection methods for incompressible flows}.
\newblock \emph{Comp. Meth. Appl. Mech. Eng.}, \textbf{195(44)}, 6011--6045.

\bibitem[{Henson and Yang(2000)}]{henson2000}
HENSON, V.E. and YANG, U.M. (2000).
\newblock \enquote{Boomer{AMG}: a parallel algebraic multigrid solver and
  preconditioner}.
\newblock \emph{Appl. Numer. Math.}, \textbf{41}, 155--177.

\bibitem[{{\lowercase{\sl hypre }}(2014)}]{hypre}
{\lowercase{\sl HYPRE }} (2014).
\newblock \emph{High Performance Preconditioners}.
\newblock Lawrence Livermore National Laboratory.
\newblock \url{http://www.llnl.gov/CASC/hypre/}.

\bibitem[{John \emph{et~al.}(2006)}]{john2006}
JOHN, V. \emph{et~al.} (2006).
\newblock \enquote{A comparison of time-discretization/linearization approaches
  for the incompressible {N}avier--{S}tokes equations}.
\newblock \emph{Comput. Meth. Appl. Mech. Eng.}, \textbf{195(44)}, 5995--6010.

\bibitem[{Kang \emph{et~al.}(2000)}]{kang2000}
KANG, M. \emph{et~al.} (2000).
\newblock \enquote{A boundary condition capturing method for multiphase
  incompressible flow}.
\newblock \emph{J. Sci. Comput.}, \textbf{15(3)}, 323--360.

\bibitem[{Liu \emph{et~al.}(1994)}]{liu1994}
LIU, X.D. \emph{et~al.} (1994).
\newblock \enquote{Weighted essentially non-oscillatory schemes}.
\newblock \emph{J. Comput. Phys.}, \textbf{115(1)}, 200--212.

\bibitem[{McCalpin(1995)}]{McCalpin1995}
MCCALPIN, J.D. (1995).
\newblock \enquote{Memory bandwidth and machine balance in current high
  performance computers}.
\newblock \emph{IEEE Computer Society Technical Committee on Computer
  Architecture (TCCA) Newsletter}, 19--25.

\bibitem[{McCalpin(2014)}]{stream}
MCCALPIN, J.D. (2014).
\newblock \enquote{{STREAM} web site}.
\newblock {\url{http://www.cs.virginia.edu/stream/}}.

\bibitem[{Moore(1965)}]{moore1965}
MOORE, G. (1965).
\newblock \enquote{Cramming more components onto integrated circuits}.
\newblock \emph{Electronics}, \textbf{38(8)}.

\bibitem[{Nethercote and Seward(2007)}]{2007valgrind}
NETHERCOTE, N. and SEWARD, J. (2007).
\newblock \enquote{Valgrind: a framework for heavyweight dynamic binary
  instrumentation}.
\newblock \emph{ACM Sigplan Notices}, vol.~42, 89--100.

\bibitem[{Nethercote \emph{et~al.}(2014)}]{valgrind}
NETHERCOTE, N. \emph{et~al.} (2014).
\newblock \enquote{Valgrind {W}eb page}.
\newblock {\url{http://valgrind.org/}}.

\bibitem[{Osher and Fedkiw(2001)}]{osher2001}
OSHER, S. and FEDKIW, R.P. (2001).
\newblock \enquote{Level set methods: An overview and some recent results}.
\newblock \emph{J. Comput. Phys.}, \textbf{169(2)}, 463 -- 502.

\bibitem[{Osher and Fedkiw(2003)}]{osher2003}
OSHER, S. and FEDKIW, R.P. (2003).
\newblock \emph{Level Set Methods and Dynamic Implicit Surfaces}.
\newblock Springer, Berlin.

\bibitem[{Osher and Sethian(1988)}]{osher1988}
OSHER, S. and SETHIAN, J.A. (1988).
\newblock \enquote{Fronts propagating with curvature-dependent speed:
  {A}lgorithms based on {H}amilton-{J}acobi formulations}.
\newblock \emph{J. Comput. Phys.}, \textbf{79(1)}, 12 -- 49.

\bibitem[{Roache(2002)}]{roache2002}
ROACHE, P.J. (2002).
\newblock \enquote{Code verification\texttrademark\ by the method of
  manufactured solutions}.
\newblock \emph{ASME J. Fluids Eng.}, \textbf{124(1)}, 4--10.

\bibitem[{Rossinelli \emph{et~al.}(2013)}]{SC2013GordonBell}
ROSSINELLI, D. \emph{et~al.} (2013).
\newblock \enquote{11 {PFLOP}/s simulations of cloud cavitation collapse}.
\newblock \emph{SC}, 3.

\bibitem[{Teigen and Munkejord(2010)}]{teigen2010a}
TEIGEN, K.E. and MUNKEJORD, S.T. (2010).
\newblock \enquote{Influence of surfactant on drop deformation in an electric
  field}.
\newblock \emph{Phys. Fluids}, \textbf{22(11)}.
\newblock Article 112104.

\bibitem[{van~der Vorst(1992)}]{vandervorst1992}
VAN~DER VORST, H. (1992).
\newblock \enquote{{Bi-CGSTAB}: A fast and smoothly converging variant of
  {Bi-CG} for the solution of nonsymmetric linear systems}.
\newblock \emph{SIAM J. Sci. Comput.}, \textbf{13(2)}, 631--644.

\bibitem[{Zhuang and Sun(2001)}]{Zhuang2001}
ZHUANG, Y. and SUN, X.H. (2001).
\newblock \enquote{A high-order fast direct solver for singular {P}oisson
  equations}.
\newblock \emph{J. Comput. Phys.}, \textbf{171(1)}, 79 -- 94.

\end{thebibliography}

\end{document}